# Fault Tolerant Dynamic Task Assignment for UAV-based Search Teams

Ali Nasir[1,2], and Mohammad AlDurgam[3,4]

[1]Interdisciplinary Research Center for Intelligent Manufacturing and Robotics, KFUPM, Dhahran, Saudi Arabia

[2]Control and Instrumentation Engineering Department, KFUPM, Dhahran, Saudi Arabia

[3]Interdisciplinary Research Center for Smart Mobility and Logistics, KFUPM, Dhahran, Saudi Arabia

[4]Industrial and Systems Engineering Department, KFUPM, Dhahran, Saudi Arabia

**Abstract**

This research offers a novel framework for dynamic task assignment for unmanned aerial vehicles (UAVs) in cooperative search settings. Notably, it incorporates post-fault UAV capabilities into job assignment techniques, assuring operational dependability in the event of sensor and actuator failures. A significant innovation is the utilization of UAV battery charge to assess range relative to search objectives, hence improving job distribution while conserving battery life. This model integrates repair, recharge, and stochastic goal recurrence, hence increasing its real-world applicability. Using stochastic dynamic programming, this method makes it simpler to determine optimal assignment policies offline so they may be implemented rapidly online. This paper emphasizes the holistic aspect of the proposed model, which connects high-level task rules to low-level control capabilities. A simulation-based case study proves its usefulness, highlighting its robustness in fault-prone and battery-variable settings. Overall, this paper proposes and demonstrates a comprehensive method for assigning UAV tasks that integrates defect awareness, battery management, and multilayer control through the use of stochastic dynamic programming.

**Keywords:** Markov decision process, multilayer control, task assignment, unmanned aerial vehicle.

## 1 Introduction

The task assignment problem for multiple unmanned aerial vehicle (UAV) missions is relevant to multiple real-life applications, e.g., surveillance, monitoring, transportation, security, and entertainment. As per the existing literature on the subject, popular considerations for task assignment problem include energy optimization, robustness, collision avoidance, and time efficiency [2]. Furthermore, there are two major types of task assignment strategies: cooperative and non-cooperative task assignments. Our proposed research focuses on the cooperative task assignment where the UAVs can be dynamically reassigned regions to be searched based on their own health and battery status of their own and that of other UAVs. The reason for focusing on cooperative task assignment is to cater for unexpected low-battery situations and other faults (e.g., sensor and actuator failures) arising in the UAVs. A unique aspect of the proposed research is to enable *feedback control-aware* task assignment, where the high-level task assignment policy models the capability of the low-level control used for motion control of a UAV.

Task assignment in general and specifically for UAV applications is a well-researched problem [2] but there are still certain issues that need attention, e.g., incorporation of uncertainties, fault tolerance, and awareness regarding reconfigurable low-level motion control. The major focus of the existing task assignment techniques proposed over the past few years has been energy optimization. Other objectives include collision avoidance, cost-efficiency, completeness, robustness, and time-efficiency. For example, the use of angle-encoded particle swarm optimization [2] for minimum-path surface monitoring with a team of UAVs has been proposed recently, where obstacle avoidance and maintenance of altitude are among the main features. In another path coverage approach, incorporation of energy constraints by a UAV [3] has been proposed, where the optimization of waypoints is achieved with Voronoi-based path generation. This scheme, however, is for an individual UAV. According to a survey on coverage path planning with a focus on UAV-based applications [4], the limited endurance of UAVs is a major concern in terms of area coverage. Also, there is a trade-off between path length and energy consumption wherever the shortest path involves sharp turns. Note that the "tasks" in a UAV-based search problem involve the allocation of specific region(s) to search through; therefore, a lot of relevant literature is found in path planning for multiple UAV missions. For example, relevant research includes energy-aware coverage path planning [5] and energy-efficient path planning [6]. Recently, the focus has also shifted from a two-dimensional to a three-dimensional path planning approach, especially for the monitoring of oil fields [7]. An integer programming-based approach for path planning of UAVs has been proposed in [8], where the two-step optimization technique is discussed, but the uncertainties involved in the problem have not been catered for. Other applications of UAV path planning include structure inspection [9], inspection of photovoltaic farms [10][11], monitoring of oil fields [12], inspection of transmission lines [13], and ensuring public safety [14]. A commonality among all of the above-mentioned contributions is that they use deterministic approaches.

Researchers have also used reinforcement learning for task assignment and path planning in multi-agent UAV-based missions. [15][16]. For example, in reference [15], Q-learning has been used to find the task allocation rules. For this purpose, a Q-network has been generated that is able to cater for uncertainty to some extent and is able to handle different tasks. An iterative approach for UAV task assignment during the path planning process has been discussed in [17], whereas a hybrid artificial potential field and ant colony optimization-based approach is presented in [18]. A Similar problem has been addressed in [19] with trajectory generation and waypoint assignment for UAVs in an urban environment. Path planning in urban environments is usually focused on obstacle avoidance [20]. We focus more on the high-level assignment of tasks and leave the detailed path planning (including obstacle avoidance) to a low-level control that is beyond the scope of this paper. A combination of partially observable Markov decision process (POMDP) and gray wolf optimization has been discussed in [21], where the gray wolf optimization is used for UAV communication systems and the POMDP is used for collision avoidance. Note that, similar to obstacle voidance, collision avoidance is also beyond the scope of this paper. Another relevant research on fault-tolerant cooperative navigation of UAVs is discussed in [22]. In this approach, the abstract faults have been considered, whereas in this paper, we focus on the specific faults that are relevant to the capabilities of the UAVs in terms of sensing and actuation. For a detailed study of the relevant issues and challenges, recent survey papers are [23], [24], and [25].

A stochastic approach for UAV search teams has rarely been discussed in the literature. References [26][27] explore the utilization of a two-level planning approach with consideration of the uncertainties involved in the area to be monitored or searched. These references proposed a two-level Markov decision process (MDP)-based approach, which is the baseline for the research being proposed here. On the other hand, there has been some utilization of MDP based models in the path planning problem. For example, [28][30] proposed an execution time reduction technique for the MDP model used in path planning in robotic systems. The reduction in execution time relies on focusing only on the accessible states inthe current state. Another MDP-based path planning scheme for mobile robots is proposed by [29]. In this scheme, ant colony optimization has been combined with the MDP to improve the efficiency of global path planning. Recently, [31] proposed an MDP-based approach for navigation on the roads. While all of the above techniques and many more similar methods have demonstrated the use of MDP for solving path planning problem, to the best of our knowledge, there has been no attempt to discuss the utilization of an MDP based approach involving fault tolerance for a team of UAVs. Therefore, in this research, we propose a two-level hierarchical MDP based approach addressing the problem of dynamic task assignment for the team of UAVs.

The proposed framework adds novelty to existing research in numerous ways. One advantage is that it is the first approach to consider post-fault UAV capabilities in terms of system properties such as controllability and observability. This data will be used to evaluate the operation's reliability in the presence of various actuator and sensor faults. Second, we propose that the state of charge of each UAV's battery be used to determine the UAV's range in relation to the search goals. In our proposed model, we use range information to help with task assignment. This method avoids deep discharge and depletion of the UAV batteries while also facilitating safe operation. In our model, we also include the ability to repair and recharge UAVs. Furthermore, we introduce recurring goals with randomly generated repeated requests to improve the reliability and security associated with real-world applications. The proposed approach optimizes dynamic task assignment using stochastic dynamic programming. Finally, the optimal policy for task assignment is computed offline, which reduces the time required for an online response.

## 2 Background and Problem Formulation

In this section, we present the background essential to comprehend the problem (e.g. definition of the state variables), followed by a formal definition of the research problem.

### 2.1 Feedback Control, Controllability, and Observability

Before discussing the proposed approach, it is important to understand certain aspects of the feedback control of UAVs. A UAV is a nonlinear dynamical system described by a state space model of the form

$$\dot{x} = f(x) + g(x)$$
$$y = h(x) \qquad (1)$$

Where $x$ is the state vector that includes states such as attitude angles, angular velocities, linear displacement, and linear velocities. The nonlinear functions $f(x)$ and $g(x)$ describe the dynamics of the UAV. The control input $u$ is designed in order to enable the UAV to track the reference trajectories with the help of onboard actuators (e.g., motor-propeller sets). The output vector $y$ denotes the measurements from the sensors mounted on the UAV (e.g., accelerometer, gyroscope, magnetometer, or inertial measurement unit).

Under normal circumstances (assuming that the UAV is equipped with a sufficient number of sensors and actuators), the UAV is controllable and observable. Controllability is the ability to move a UAV from any initial state $x = x_0$ to any final state $x = x_f$ in finite time and using (piecewise continuous) finite control input. In contrast, observability is the ability to determine the initial state (and hence the state vector for all time) of a UAV based on the sensor measurements and the knowledge of the control input $u$. For nonlinear systems such as a quadrotor UAV, the controllability and observability properties can be determined locally (in the vicinity of an equilibrium point $x^*$) using the well-known controllability and observability matrices. For example, the controllability matrix is given by

$$Q = [B \quad AB \quad A^2B \ldots \quad A^{n-1}B \quad A^nB] \quad (2)$$

Where $A = \frac{\partial f}{\partial x}|_{x=x^*}$ and $B = \frac{\partial g}{\partial x}|_{x=x^*}$ are the Jacobian matrices corresponding to the nonlinear functions $f(x)$ and $g(x)$ respectively, that are evaluated at the equilibrium point $x = x^*$. The quadrotor is deemed controllable (locally in the vicinity of the equilibrium point) if the rank of $Q$ matrix is equal to the size of the state vector $x$. Similarly, the observability is determined using the matrix.

$$O = [C \quad CA \quad CA^2 \ldots \quad CA^{n-1} \quad CA^n]^T \quad (3)$$

Where $C = \frac{\partial h}{\partial x}|_{x=x^*}$ and the observability is guaranteed in the vicinity of the equilibrium point if the rank of $O$ matrix is equal to the size of the state vector.

Note that the controllability of a UAV depends upon the availability and proper working of the actuators, whereas the observability depends upon the availability and proper working of the sensors. If one or more actuators fail, the UAV may lose controllability (while it may still be stabilized, i.e., uncontrollable states are asymptotically stable). Similarly, if one or more of the sensors fail, the UAV may lose observability (while it may still be detectable, i.e., unobservable states are asymptotically stable). Hence, there are nine possible discrete situations with respect to combinations of the observability and controllability properties of a UAV, as shown in Table 1. We shall be considering these situations in our model, which is discussed in the next section.

Table 1: Fault states of a UAV

| Index | State | Ability of the feedback control |
|---|---|---|
| 1 | controllable and observable | high |
| 2 | controllable and detectable | medium |
| 3 | stabilizable and observable | medium |
| 4 | stabilizable and detectable | medium |
| 5 | controllable and undetectable | low |
| 6 | stabilizable and undetectable | low |
| 7 | unstabilizable and observable | low |
| 8 | unstabilizable and detectable | low |
| 9 | unstabilizable and undetectable | low |

## 2.2 Power Consumption and Range

UAVs that we consider in this work use electric power supplied by a battery. The flight time of a UAV is usually limited to less than an hour (for a quadcopter). The main source of power consumption is the motors that drive the propellers. Secondly, the payload equipment, such as a camera and onboard data processing electronics, consumes power. Hence, the range of a UAV is limited. In this paper, we consider the limited range constraint in terms of the feasible set of task assignments given the location of the UAV and the state of charge of its battery. Suppose that the power consumption of the UAV is depicted by

$$p_{UAV} = p_m + p_p + p_e \quad (4)$$

Were $p_{UAV}$ is the total power consumed by the UAV at an instant, $p_m$ is the power consumed by the motors, $p_p$ is the power consumed by the payload equipment, and $p_e$ is the power consumed by the onboard electronics. Then the flight duration is calculated as follows:

$$t \leq \frac{SOC \times bc \times v}{p_{UAV}} \quad (5)$$

Where $SOC$ is the normalized state of charge for the UAV battery that ranges between 0 and 1, $bc$ is the full battery capacity in terms of ampere-seconds, $v$ is the nominal voltage across the terminals of the battery, $t$ is time (in seconds), and $fd$ is the maximum flight duration of the UAV (in seconds). Note that the maximum flight duration is given by

$$fd_{max} = \frac{bc \times v}{p_{UAV}} \quad (6)$$

Now the above flight duration depends upon the value of $p_{UAV}$ which is not a constant in general. Hence, the maximum range can be approximated from the flight duration by assuming a minimum value of $p_{UAV}$ that is required for one dimensional motion of the UAV at a constant altitude.

From the flight duration, the range of the UAV is calculated by assuming an average speed $\delta$ (meters per second) as

$$r = t \times \delta \quad (7)$$

Based on this range, and the distance required to be traveled against various task assignments, the feasibility of the task assignment is calculated as

$$fsb_\alpha = \begin{cases} 1 & if\ r \geq d_\alpha(l) \\ 0 & otherwise \end{cases} \quad (8)$$

Here, $fsb_\alpha$ is the feasibility flag for an assignment $\alpha$, and $d_\alpha(l)$ is the distance to be traveled in order to complete the assignment $\alpha$ from the starting location $l$ of the UAV. Note that, $d_\alpha$ is calculated based on the sum of distances between the waypoints to be visited in the assignment $\alpha$ in such a way that the waypoints are arranged in a sequence that results in the minimum sum of distances. Hence, a nontrivial amount of calculation is involved in the determination of $d_\alpha(l)$.

## 2.3 Decision-Making under Uncertainty and MDP

One of the most popular frameworks that facilitates the calculation of an optimal policy in the presence of uncertainty is an MDP. MDP is a way of modeling a complex decision-making problem that allows for the calculation of optimal policy using stochastic dynamic programming algorithms such as policy iteration and value iteration. In order to create a mathematical model as an MDP, one needs to specify five items, i.e., a set of states, a set of actions, a state and action dependent cost or reward function, a state and action dependent transition probability tensor, and a discount factor whose value is selected from the interval (0,1). Hence, an MDP model is a five-tuple,

$$MDP \leftarrow \{S, D, J, P, \gamma\} \quad (9)$$

Where the set of states ($S$) represents an instantiation of underlying discrete state variables. The set of decisions (or actions), $D$, includes possible actions that can be executed in order to make progress from the current state towards a goal state. The cost (or reward) function $J$ differentiates between the desirable and undesirable states and it can also be used for indicating the cost associated with the decisions (or actions). The probability tensor $P$ holds the state transition probabilities in the form $\Pr(s'|s, d)$, i.e., the probability of reaching a state $s' \in S$ from a state $s \in S$ by executing an action $d \in D$. Finally, the discount factor ($\gamma$) defines the depreciation in the value of states in the long-term. If the value of $\gamma$ is selected to be close to one, then a state will have a similar value in the future as it has now. On the other hand, if $\gamma$ is chosen to be close to zero, then a state in the future has a significantly lower value than its value in the present. Usually, a higher value of $\gamma$ is selected when long-sighted policy is desirable.

Other than the five elements mentioned above, an MDP also needs a decision-making horizon to be specified. We use the infinite horizon in this paper because our task assignment problem is of a continuous nature without a pre-specified end. Secondly, we need to know the decision epoch, i.e., the duration of time between two consecutive decisions. In our case, the decision epoch shall be based on the time required for a UAV to move from one cell (region) of the search grid (area) to an adjacent one (see Figure 1).

Once a decision-making problem is modeled as an MDP, it can be solved using any of the stochastic dynamic programming algorithms, such as policy iteration or value iteration. Value iteration is preferred in this paper because we shall be using the optimal values of the local MDP states in the calculation of the global decision (this will be further explained in the next section). The value iteration algorithm computes the optimal value for each state by using Bellman's equation as follows:

$$V^{t+1}(s) = \max_{d \in D} \left\{ -J(s,d) + \sum_{s' \in S} \gamma P(s'|s,d) V^t(s') \right\} \quad (10)$$

In order to start the value iteration as shown above, the values are first initialized to an arbitrary value. After a certain number of iterations (regardless of the initial value of the states), the values of the states converge, i.e., $V^{t+1}(s) \approx V^t(s), \forall s \in S$. The iterative process can be stopped using a stopping criterion of the form:

$$\|V^{t+1} - V^t\|_\infty < \eta \quad (11)$$

Where $V^t$ is the vector of the values of all the states at time $t$. When the above criterion is met, the corresponding values of the states are called the optimal values, denoted as $V^*(s), \forall s \in S$. Using the optimal values, the optimal policy is calculated using the following equation:

$$\pi(s) = \underset{d \in D}{\operatorname{argmax}} \left\{ -J(s,d) + \sum_{s' \in S} \gamma P(s'|s,d) V^*(s') \right\} \quad (12)$$

Note that, $\pi(s) \in D$, i.e., $\pi: S \to D$.

### 2.4 Problem Setup and Formulation

The layout of the problem is shown in Figure 1. We assume that the operation of the team of UAVs is to be performed within a bounded area that is divided into regions (the regions in the figure are shown to be identical in size, but in practical situations, slight variations in the size of the region are acceptable). The problem is that of assigning different regions (waypoints) to each UAV given the externally controlled search priorities. We assume that there is a base station (or ground control station) for the UAVs. Furthermore, the base and the UAVs can communicate with each other using a wireless link. More assumptions regarding the problem are stated as follows:

- Each UAV can be assigned or reassigned the task of monitoring various regions. Tasks are assigned by the base station.

- The controllability and observability analysis have been performed offline for cases of onboard sensor and actuator faults, i.e., under any fault condition, the controllability and observability of a UAV are known.

- Each UAV is equipped with fault-tolerant control, with appropriate control law switching is in place against actuator and sensor faults.

- Measurement of the state of charge (SoC) of the UAV battery is available.
- Each UAV is mounted with a camera (payload sensor), and the health status of the camera is available.

The general problem of dynamical task assignment involves three elements, i.e., waypoint assignment, waypoint sequencing, and feedback control. In this paper, we focus only on the waypoint assignment. The sequencing and feedback control have been assumed to be in place.

Dynamic task assignment needs to be optimized in order to save time and energy. Also, there are some uncertainties involved in the problem, e.g., the occurrence of faults in the UAV sensors and actuators, camera failure, and changes in search region priorities. Therefore, we must model this problem in a way that allows us to perform stochastic optimization.

MDP serves as a stochastic dynamic programming methodology, valuable for making decisions under uncertainty [32]. Its broad array of practical applications and effective solution techniques have been well-documented [33] [32].

According to [34], key merits of MDP include: 1) its capacity to derive optimal policies for both limited and infinite planning durations, the latter being suited for modeling stationary systems, 2) MDP accommodates diverse cost criteria that may rely on the initial system state, subsequent system state, or both [32], and 3) the real-world deployment of MDP in decision-making under uncertainty has been extensively observed [33].

In order to be able to use stochastic dynamic programming, the decision-making problem is be formulated as a Markov decision process. In the next section, we present a proposed solution for the problem and associated mathematical models.

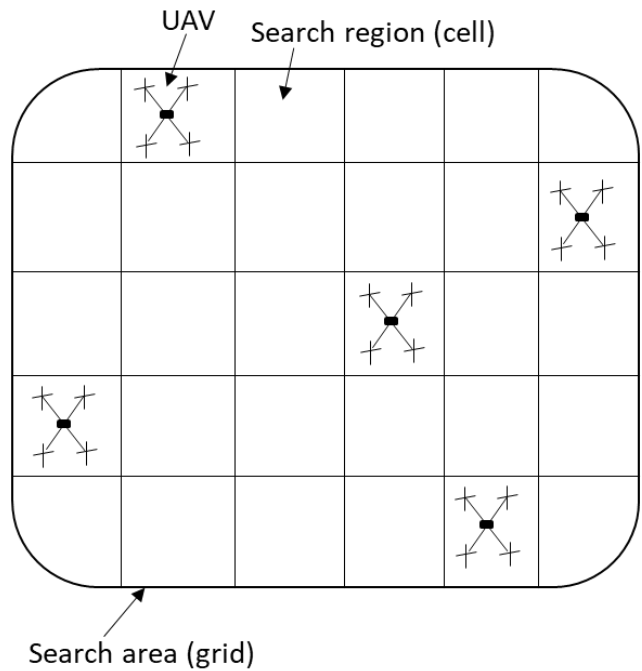

Figure 1: Layout of the search area under assumption

# 3 Proposed Solution Architecture and Mathematical Model

The overview of the proposed solution for the dynamical task assignment problem is shown in Figure 2. Our proposed solution uses a combination of centralized and decentralized computations. Each UAV has its own decision-making model that is used to calculate the value (a measure of cost) associated with each search task for the individual UAV. All UAVs communicate the results of calculations, i.e., the values associated with the search tasks to the base station. The base station also receives the search task priorities from the external users (which change dynamically depending upon the user preferences). After collecting all the information from the UAVs and the users, the base station uses stochastic optimization to assign the search tasks to the UAVs. The base station also communicates the updated search region priorities to the UAVs that is used by the UAVs to update the value associated with each search task. This cycle repeats indefinitely.

Next, we define our proposed decision-making models for the UAVs and the base station.

## 3.1 MDP Model for Fault-Aware Goal Bidding at UAV Level

In this section, we present the MDP model for fault-aware goal bidding at each UAV. The reason for modeling local goal-bidding as an MDP is to facilitate optimization in the presence of uncertainty. Specifically, we discuss the set of states, available decision variables, reward criteria, and transition probabilities involved in the goal-bidding problem.

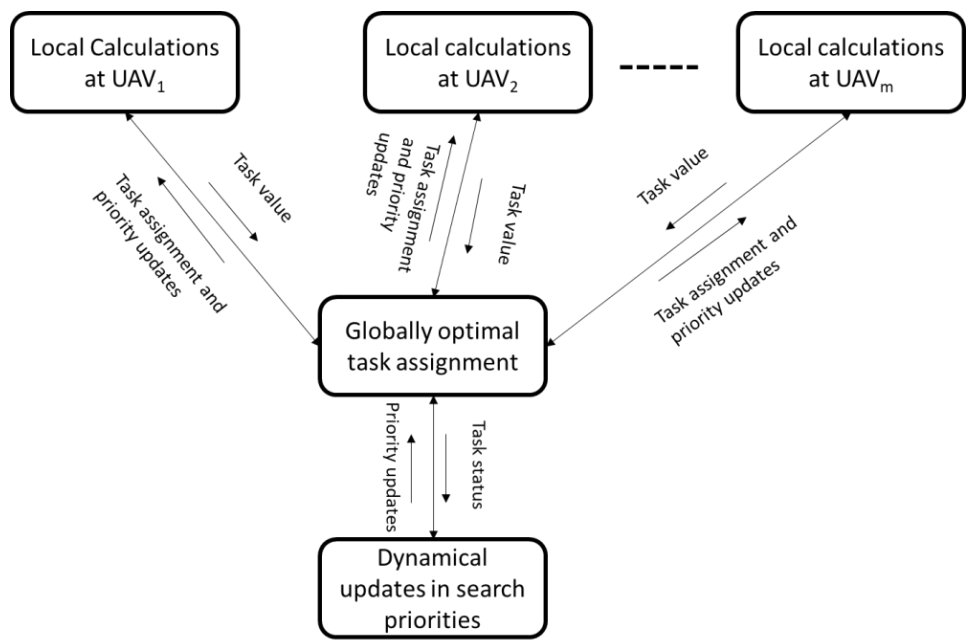

*Figure 2: Proposed Solution Architecture*

### 3.1.1 State Variables

In order to correctly determine the bidding value corresponding to a goal (or search task), the UAV must account for its state of charge (to determine which of the tasks are within range), the occurrence of faults in the feedback control-related or payload-related components, the location of the UAV, and the priorities among the tasks. Consequently, the set of states is defined as,

$$\begin{aligned}
&S = \{s_1, s_2, \ldots, s_n\}, \\
&s_i = \{f_i, R_i, G_i, l_i, c_i\}, i \in \{1, 2, \ldots, n\}, \\
&f_i \in \{1, 2, \ldots, 18\}, \\
&R_i = \{r_{1,i}, r_{2,i}, \ldots, r_{k,i}\}, r_{j,i} \in \{0,1\}, j \in \{1, 2, \ldots, k\}, i \in \{1, 2, \ldots, n\}, \\
&G_i = \{g_{1,i}, g_{2,i}, \ldots, g_{k,i}\}, g_{j,i} \in \{0,1,2\}, j \in \{1, 2, \ldots, k\}, i \in \{1, 2, \ldots, n\}, \\
&l_i \in \{1, 2, \ldots, q\}, \\
&c_i \in \{0, 1, 2, \ldots, k\}
\end{aligned} \qquad (13)$$

The above equation defines a total of $n$ states in the set of states for the MDP model. Each state ($s_i$) is composed of five types of variables. The first variable is a scalar ($f_i$) representing the health status of the UAV. The values for this variable range from 1 to 18, based on the discussion in Section 2.1. Recall that there are nine possible states of UAV in terms of controllability, observability, stabilizability, and detectability, as shown in Table 1, and since there is a possibility of payload camera failure as well, the total number of possible fault state values becomes 18. The second variable ($R_i$) includes $k$ binary flags (where $k$ is the number of goals). Each binary flag ($r_{j,i}$) in $R_i$ represents if the UAV has enough energy to attempt the $j^{th}$ goal ($r_{j,i} = 1$) or not ($r_{j,i} = 0$). The binary flags ($r_{j,i}$) in the state space use the underlying calculations regarding the range of UAV discussed in Section 2.2. The third variable in the state space ($G_i$) includes $k$ goal priority flags ($g_{j,i}$). Each goal priority flag has been indicated in the equation to have three possible values representing low priority ($g_{j,i} = 1$), high priority ($g_{j,i} = 2$), no priority ($g_{j,i} = 0$). No priority means that either the goal has already been achieved or is not desirable by the user at the moment. The fourth variable in the state space ($l_i$) is the location of the UAV. The search area has been divided into $q$ search regions (see Figure 1). The role of location is to determine the cost of execution of a search task, as will be clear when the reward function is defined. Finally, the fifth variable indicates the active commitment of the UAV with any of the goals, ($c_i > 0$) or lack of commitment ($c_i = 0$).

Note that we have gathered in the state variables all the important information that is required for the determination of the bid value for a goal from the perspective of a single UAV. Given the diversity of the information involved and the fact that some of the variables are random in nature, e.g., the occurrence of a fault or the priority level of a goal or a region, it is not feasible to solve this problem using conventional task assignment techniques such as existing solutions of the travelling salesman problem or its variants. Therefore, we have selected the MDP as our modeling tool. Next, we describe the set of available decisions for the base station.

### 3.1.2 Decision Variables

At any given time, the UAV can decide to pursue a goal (search task), request a recharge, or request service (replacement of faulty equipment). The decision-making is event-based, i.e., a decision is made by the UAV only if it moves from one state to another (different) state. The time duration involved in the transition from one state to another is variable. This flexibility does not affect the MDP model and associated calculation of the state value function because the time is not modeled directly in the MDP. Mathematically, the set of decisions is given by

$$D = \{d_1, d_2, \ldots, d_k, serv, charge, Continue\} \qquad (14)$$

As shown above, there are four types of decisions in the model. There are $k$ decisions regarding the commitment to a search task (or goal), a decision to request service ($serv$), a decision to request recharging (or replacement) of the batteries ($charge$), and a decision to keep pursuing the currently assigned goal ($Continue$).

### 3.1.3 The Cost Criteria

The cost criteria for the MDP model involve four aspects. One is the reward associated with the completion of a search task (goal). Second is the cost associated with execution of the search task (energy cost). Third is occurrence of faults. The fourth aspect of the cost is the inability to attempt search tasks due to a lack of available range. Hence, the cost function depends on the state and the decision variables. In order to avoid negative values and zero crossing, the reward and cost have been consolidated into a single function given by

$$J(s_i, d) = \sum_{j=1}^{k} \eta_j g_{j,i} r_{j,i} \left(1 - \mathbb{I}_j(c_i)\right) + \hbar(d, l_i) + f(f_i, r_i) + \sum_{j=1}^{k} \delta_j g_{j,i} (1 - r_{j,i}) \quad (15)$$

The above equation consists of four terms. The first term penalizes the unassigned goals that are within range. Here, $\eta_j$ are user-defined constants for specifying the importance of each goal. The second term is an abstract function of the location of the UAV and the assigned task. This function can be defined by using the Euclidian distance between the UAV and the location of the goal that is being assigned to the UAV. The third term represents the cost of having faults and the cost of having a low battery. As example of the definition of the functions $\hbar(d, l_i)$ and $f(f_i, r_i)$ are presented in the case study section. The final term in the cost function is the specific cost of not being able to reach the goal locations due to low battery. Here the constants $\delta_j$ are user-defined.

### 3.1.4 Transition Probabilities

Out of the four types of state variables discussed above, the two that involve randomness are the fault variable ($f_i$) and the goal priority variable ($G_i$). Hence, the state transition probabilities involve the joint probability distribution of the fault and goal variables. Consequently, the general form of the state transition probability function is given by

$$P(s_i, d, s_j) = \Pr(s_j | s_i, d) = \Pr(f_j = f, g_{1,j} = g_1, \ldots, g_{k,j} = g_k | f_i = \bar{f}, g_{1,i} = \bar{g}_1, \ldots, g_{k,i} = \bar{g}_k, d) \quad (16)$$

The joint probability distribution of the faults and the goal priorities is required for solving the MDP model. we assume that the fault probabilities are independent of the goal priority probabilities. Furthermore, the change in goal priorities is independent of the decision variable. Therefore, we can rewrite the state transition probabilities as

$$P(s_i, d, s_j) = \Pr(g_{1,j} = g_1, \ldots, g_{k,j} = g_k | g_{1,i} = \bar{g}_1, \ldots, g_{k,i} = \bar{g}_k, d) \Pr(f_j = f | f_i = \bar{f}, d) \quad (17)$$

The above conditional probability formulation may be further simplified by assuming conditional independence among the search goals, i.e., the priority level of one goal being independent of the priority level of the other goals. Initial probabilities can be defined based on experimental data or on experience. In such cases, the probability values can be improved using online machine learning, but such a discussion is an avenue for future research.

### 3.1.5 Calculation of task values

The task values are calculated using the value iteration algorithm, where initially, the value of each state may be set to zero, and then the values are updated iteratively using equation (10) discussed earlier. The convergence of the values is tested using the criterion in equation (11) with $\eta$ being a number smaller than $10^{-6}$. Now, the final step (i.e., policy calculation) in the value iteration algorithm is altered in this case. Instead of finding the best action for each state (as in equation 12), the value corresponding to each goal assignment action is generated using the following equation:

$$\mathcal{V}(s,d) = -J(s,d) + \sum_{s' \in S} \gamma P(s'|s,d)V^*(s') \quad (18)$$

Note that the above equation generates a value for each state-action pair. The step-by-step process for the calculation of task values is shown in Figure 3. These values are used in the MDP model for fault tolerant dynamic task assignment, as discussed in the next section.

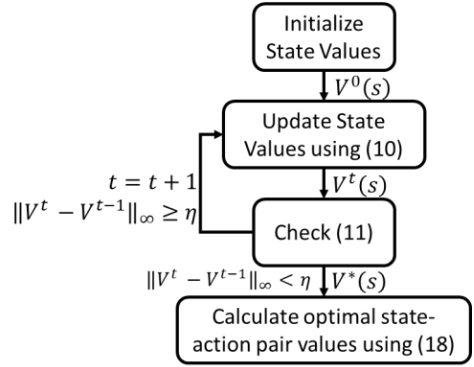

Figure 3: Calculation of task values

## 3.2 MDP Model for Fault Tolerant Dynamic Task Assignment

In this section, we present the stochastic decision-making model for dynamic task assignment.

### 3.2.1 State variables

Dynamic task assignment requires information from the UAVs as well as from the user in order to make optimal decisions in the presence of uncertainty. Specifically, the user should provide updated search goal priorities. Each UAV shall provide values associated with each search goal. This value is indicative of the locally optimized assessment of the UAV. Furthermore, central decision-making also requires the current status of each UAV (i.e., the engagement of the UAV in a task, recharging, or service). Finally, since the tasks are being assigned (and reassigned) dynamically, the decision-making process must be aware of the most recent task assignment. Consequently, the set of states for central decision-making is given by

$$\begin{aligned}
\mathcal{S} &= \{s_1, s_2, \dots, s_N\} \\
s_i &= \{G_i, \mathcal{A}_i, F_i, \mathcal{V}_i, \mathcal{D}_i\} \\
G_i &= \{g_{1,i}, g_{2,i}, \dots, g_{k,i}\},\ g_{j,i} \in \{0,1,2\} \\
\mathcal{A}_i &= \{a_{1,i}, a_{2,i}, \dots, a_{z,i}\},\ a_{j,i} \in \{0,1,2,\dots,k\} \\
F_i &= \{f_{1,i}, f_{2,i}, \dots, f_{z,i}\},\ f_{j,i} \in \{1,2,\dots,18\} \\
\mathcal{V}_i &= \{v_{1,1,i}, v_{2,1,i}, \dots, v_{k,z,i}\},\ v_{j,l,i} \in \{0, \delta, 2\delta, \dots, \delta_{max}\} \\
\mathcal{D}_i &= \{d_{1,i}, d_{2,i}, \dots, d_{z,i}\} \\
d_{j,i} &\in \{0,1\}
\end{aligned} \quad (19)$$

The above equation indicates that there are five variables in the state space. Two of the variables (i.e., $G_i$ and $f_i$) are the same as the ones in the state space for the UAV bidding model (Section 3.1). Note that the central decision-making process is aware of the faults of all UAVs ($z$ is the total number of UAVs). The variable $\mathcal{A}_i$ is a vector of $z$ variables (where $z$ is the total number of UAVs) of the form $a_{j,i}$ indicating the assignment of $j^{th}$ UAV in $i^{th}$ state. The assignment of each UAV could have one of the $k+1$ values, where each nonzero value represents the search goal to be pursued by the UAV and the zero value indicates that the UAV is currently assigned no search task. Next, $\mathcal{V}_i$ include value variables associated with each goal-UAV pair. The generation of value has been discussed in the previous section. Finally, $\mathcal{D}_i$ represents the availability flag for each UAV. UAVs shall be unavailable during the service or recharging time. Note that each task assignment involves assigning of one of the $k$ goals to each of the $z$ UAVs. In the case of $k < z$, some of the UAVs shall be assigned no task. Whereas in the case of $k > z$, some of the goals shall remain unassigned. The details of the closed-loop execution will be discussed later.

### 3.2.2 Decision variables

The set of decisions for the MDP model is straight-forward and is represented as

$$M = \{\mu_1, \mu_2, \dots, \mu_x\} \tag{20}$$

In the above equations, $x$ is the total number of ways in which $k + 1$ tasks (including $k$ goal assignments and one possibility of no assignment) can be assigned to $z$ UAVs.

### 3.2.3 Cost function

The cost associated with the task assignment must encourage swift achievement of high-priority goals, the safety of the UAVs, conservation of their energy, and overall reliability of the operation. Consequently, we propose the following cost function formulation:

$$\mathcal{J}(s_i) = \sum_{j=1}^{k} \zeta_j g_{j,i} e^{g_{j,i}} + \mathcal{H}(F_i, \mathcal{A}_i, \mathcal{V}_i, \mathcal{D}_i) \tag{21}$$

The above cost function includes two terms. The first term represents the cost associated with the goals that have not been achieved. Here, $\zeta_j$ are user defined constant parameters. The second term is an abstract function that is used to specify the cost associated with the unavailability of UAV and the cost associated with neglecting the UAV preferences in task assignment. An example of the specification of the function $\mathcal{H}(d_i, F_i, \mathcal{A}_i, \mathcal{V}_i)$ is provided in the case study section.

### 3.2.4 Transition probabilities

The state transition probabilities for central decision-making are affected by two types of random variables: goal priorities and UAV faults. As a result, we can mathematically represent the state transition probabilities as

$$\mathcal{P}(s_j|s_i, d) = \Pr(F_j, G_j|F_i, G_i, d) \tag{22}$$

For the case where the occurrences of faults are independent of the changes in goal priorities, we can rewrite the above equation as

$$\mathcal{P}(s_j|s_i, d) = \Pr(F_j|F_i, d) \Pr(G_j|G_i, d) \tag{23}$$

Furthermore, independence may be established between each of the variables, with $F$ and $G$ resulting in the following expression:

$$\mathcal{P}(s_j|s_i, d) = \prod_{l=1}^{z} \Pr(f_{l,j}|f_{l,i}, d) \prod_{o=1}^{k} \Pr(g_{o,j}|g_{o,i}, d) \tag{24}$$

The above simplification of conditional probabilities makes computations easier. Also, keep in mind that random variables are used in both UAV bidding and dynamic task assignment. As a result, the data required to specify or calculate the conditional probabilities is the same in both models.

## 3.3 Closed loop implementation

The closed loop implementation requires pre-calculation of the optimal bidding for each UAV for all possible states and pre-calculation of the optimal dynamical task assignment policy. This is an advantage of using MDP-based decision-making, i.e., Because most calculations are pre-performed, online decision-making is faster. Figure 4 presents the closed-loop execution algorithm for the proposed approach. First, the updated goal preferences are obtained from the user. These values are stored in the variables $g_{1,i}, g_{2,i}, \ldots, g_{k,i}$. Second, the sensor values and current assignment of each UAV are collected (each UAV collects data from its own sensors). Third, the states of the UAVs are updated (note that the optimal values corresponding to each state have been pre-computed). Fourth, the task preferences of each UAV are communicated to the base station (wireless communication is involved in this step). Finally, the state of the dynamic task assignment is updated and the optimal decision rule (pre-computed) is executed. Consequently, the task assignments for each UAV are updated and communicated to the UAVs using wireless communication link. Figure 5 shows the block diagram for the closed loop implementation of the proposed approach. As discussed in the algorithm, each cycle of the decision-making process begins with updating the search goal priorities (these priorities are determined by the user). Based on the updated priorities and the pre-processed sensor readings from the UAVs, the optimal values (and hence the bids for the search goals) are determined and communicated to the dynamic task assignment policy. The task assignment policy then assigns search tasks to the UAVs. Note that the duration of the decision-making cycle is not specified here. This is because MDP-based decision-making (as opposed to differential equations-based control) allows for the event-based execution. For example, in this case, the events include the user changing any of the existing goal priorities, the completion of a search goal, or the occurrence of a fault in a UAV.

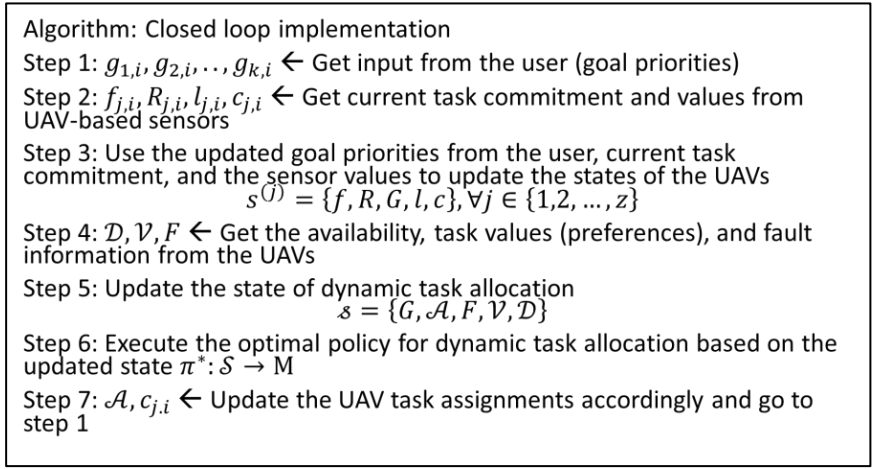

Figure 4: Closed loop implementation algorithm

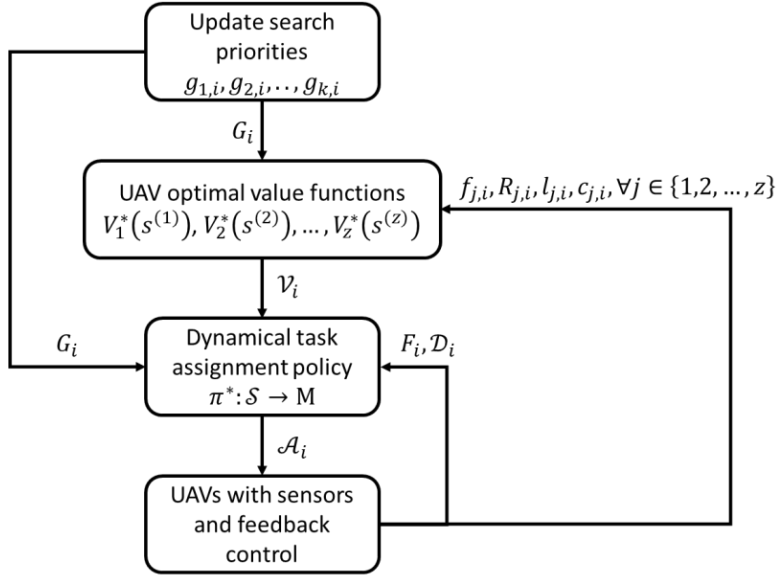

*Figure 5: Closed loop implementation of dynamical task assignment*

## 4 Numerical Case Study

In this section, we perform a detailed analysis of the proposed framework with the help of a numerical simulation-based case study. In our case study, we have used the layout shown in Figure 6 with three search goals, two UAVs, and eight search regions (locations). The details of the parameter values are presented in the next subsection, followed by discussion on the closed loop execution results and the key trends in the optimal task assignment policy calculated using the proposed framework.

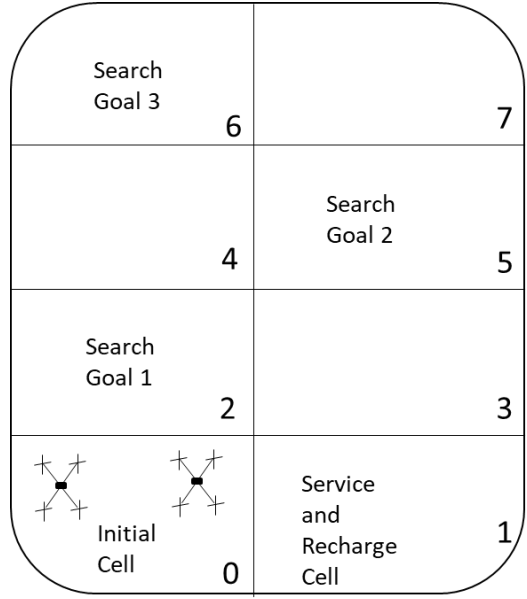

*Figure 6: Search grid layout for the case study*

## 4.1 Parameter values

The parameter values are of three types: cost function parameters, transition probability parameters, and problem size parameters. All of the parameter values, including the definitions of the abstract functions, are provided in Table 2. The selection of the parameter values is based on rough estimates of the costs associated with each of the mission activities. Nevertheless, these parameter values are for demonstration purposes only and should not be blindly adopted in real-life missions. Each multiple UAV-based task assignment problem shall have its own cost parameter values based on the purpose of the search, the cost of the UAVs utilized in the mission, battery recharging cost, UAV repair cost, and the UAV energy consumption rate.

*Table 2: Parameter values for the case study*

| Parameter | Value | Remarks |
|---|---|---|
| $n$ | 124,416 | number of states in UAV-based decision-making (eq. 13) |
| $k$ | 3 | number of goals (eq. 13) |
| $q$ | 8 | number of search regions (eq. 13) |
| $\eta_1, \eta_2, \eta_3$ | 50,70,100 | cost parameters (eq. 15) |
| $\delta_1, \delta_2, \delta_3$ | 50,70,100 | cost parameters (eq. 15) |
| $\hbar(d, l_i)$ | $\hbar(d, l_i) = \begin{cases} \begin{cases} 0 & if\ l_i = 2 \\ 2 & if\ l_i > 5 \\ 1 & otherwise \end{cases} & if\ d = 1 \\ \begin{cases} 0 & if\ l_i = 5 \\ 2 & if\ l_i < 2 \\ 1 & otherwise \end{cases} & if\ d = 2 \\ \begin{cases} 0 & if\ l_i = 6 \\ 2 & if\ l_i < 4 \\ 1 & otherwise \end{cases} & if\ d = 3 \end{cases}$ | abstract function for search cost (eq. 15) |
| $f(f_i, r_i)$ | $f(f_i, r_i) = \begin{cases} 500(r_1 + r_2 + r_3) & if\ f > 9 \\ 200(r_1 + r_2 + r_3) & if\ 4 < f < 10 \\ 50(r_1 + r_2 + r_3) & otherwise \end{cases}$ | abstract function for search cost (eq. 15) |
| $\Pr(f_j \in \{2,3,4\}\|f_i = 1)$ | 0.1 | fault probability (eq. 17) |
| $\Pr(f_j > 4\|f_i \in \{2,3,4\})$ | 0.4 | fault probability (eq. 17) |
| $\Pr(f_j > 9\|f_i \in \{2,3,4\})$ | 0.6 | fault probability (eq. 17) |
| $\Pr(f_j > 9\|f_i > 9)$ | 1 | fault probability (eq. 17) |
| $\gamma$ | 0.95 | discount factor (eq. 18) |
| $N$ | 559,872 | number of states in the dynamic task assignment (eq. 19) |
| $x$ | 12 | number of assignment choices for task assignment (eq. 20) |
| $\zeta_1 = \zeta_2 = \zeta_3$ | 100 | cost parameters (eq. 21) |
| $\mathcal{H}(F_i, \mathcal{A}_i, \mathcal{V}_i, \mathcal{D}_i)$ | $h_1(F_i, \mathcal{A}_i) + h_2(\mathcal{A}_i, \mathcal{V}_i) + h_3(\mathcal{A}_i, \mathcal{D}_i)$ | abstract cost function (eq. 21). Here, the function argmax2 finds the value of |
| | $h_1(F_i, \mathcal{A}_i)$ $= \begin{cases} 50(a_{j,i} \neq 0) & \text{if } f_{j,i} \in \{2,3,4\} \\ 100(a_{j,i} \neq 0) & otherwise \end{cases}$ $h_2(\mathcal{A}_i, \mathcal{V}_i) = \begin{cases} 0 & \text{if } a_{j,i} = \underset{l \in \{1,\dots,k\}}{\mathrm{argmax}}\, v_{l,j,i} \\ 1 & a_{j,i} = \underset{l \in \{1,2,\dots,k\}}{\mathrm{argmax2}}\, v_{l,j,i} \\ 2 & otherwise \end{cases}$ $h_3(\mathcal{A}_i, \mathcal{D}_i) = 20(a_{j,i} \neq 0)(1 - d_{j,i})$ | the variable that corresponds to second-highest value of the function. |
| $\Pr(g_j = 0 \vert g_i \neq 0)$ | 0.9 | goal achievement probability for healthy UAV that is assigned to the goal (eq. 24) |
| $\Pr(g_j = 0 \vert g_i \neq 0)$ | 0.2 | goal achievement probability for partially faulty UAV that is assigned to the goal (eq. 24) |

## 4.2 Sample Trajectories

In this subsection, we discuss the results of using the optimal task assignment policy with different initial conditions. Specifically, we show how the team of UAVs is assigned tasks in normal situations, post-fault situations, and in a low battery situation. The chain of events and policy decisions indicate that our proposed method works well despite the occurrence of faults and in situations where a UAV is running low on battery. Table 3 presents the initial values of the state variables for each of the three sample cases studied in this section. Case 1 presents the outcome of using our proposed framework in a normal situation. Case 2 investigates the results under severe fault conditions for one of the UAVs. Finally, case 3 presents the situation where one of the UAVs is running low on battery. Figure 7 and Figure 8 show the results obtained from Case1. Note that under normal situation, the decision-making policy is able to achieve all of the tree goals with two sets of assignments, i.e., [2, 1] and [3, 2]. Here [2, 1] means that UAV 1 is assigned search goal 2 and UAV 2 is assigned search goal 1. Later on, UAV 1 is assigned with search goal 3 and UAV 2 is assigned with search goal 2. As all three goals are achieved, the task assignment for both UAVs settles at [0, 0]. Here, it is important to note that the outcome of a decision taken at decision epoch $i$ arrives at decision epoch $i + 1$, i.e., in the next decision epoch.

Figure 9 and Figure 10 present the results for case 2. In this case, UAV 2 has a severe fault. Therefore, the task assignment sequence is [1, 0], [2, 0], and [3, 2] before it settles to [0, 0]. Note that due to the initial unavailability of UAV 2, goals 1 and 2 have been assigned to UAV 1, and when UAV 2 becomes available in the third decision epoch, the task assignment is [3, 2]. The assignment of goal 2 for the second time to UAV 2 (after it has been previously assigned to UAV 1) serves two purposes. First, it ensures that goal 2 is achieved with more certainty, second, it places UAV 2 at a location where all three goals are easily accessible (for the case that any of the goals is re-initiated by the user in the future).

The results from Case 3 are shown in Figure 11 and Figure 12. Here, UAV 1 is initially unavailable due to low battery but becomes available after one decision epoch. The goals are achieved using the task assignment sequences [0, 1] and [3, 2].

*Table 3: Initial States of Trajectory Cases*

| Case ID | Goal Status | UAV Commitment | UAV Availability | UAV Battery | UAV Faults | UAV Preferences |
|---|---|---|---|---|---|---|
| 1: Normal | $g_1$: High priority  $g_2$: High priority  $g_3$: Low priority | $u_1$: None  $u_2$: None | $a_1$: Available  $a_2$: Available | $R_1$: full  $R_2$: full | $f_1$: No fault  $f_2$: No fault | $v_1$: None  $v_2$: None |
| 2: Faulty UAV | $g_1$: High priority  $g_2$: High priority  $g_3$: Low priority | $u_1$: None  $u_2$: None | $a_1$: Available  $a_2$: Available | $R_1$: full  $R_2$: full | $f_1$: No fault  $f_2$: Severe fault | $v_1$: None  $v_2$: None |
| 3: Low Battery | $g_1$: High priority  $g_2$: High priority  $g_3$: Low priority | $u_1$: None  $u_2$: None | $a_1$: Available  $a_2$: Available | $R_1$: low  $R_2$: full | $f_1$: No fault  $f_2$: No fault | $v_1$: None  $v_2$: None |

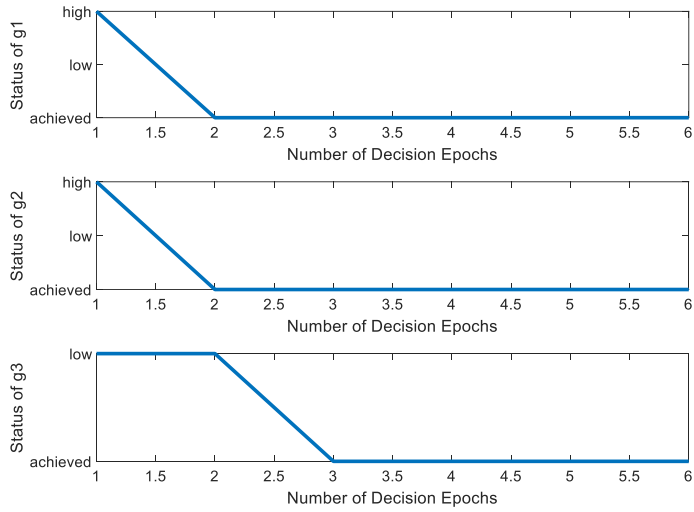

*Figure 7: Status of the goal flags for case 1*

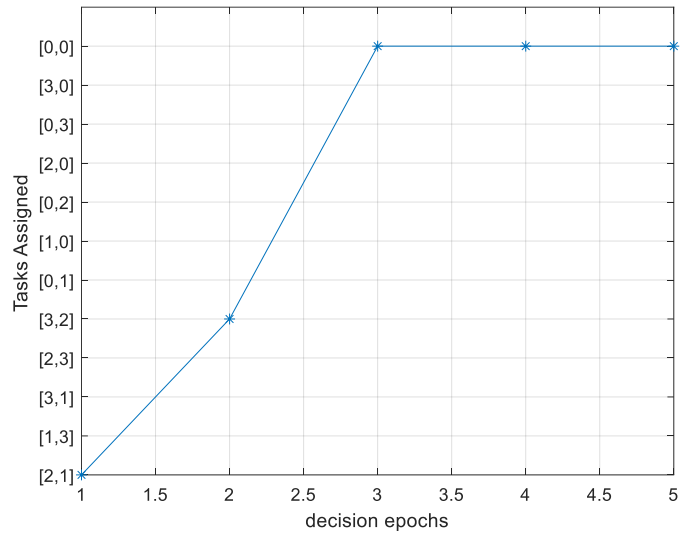

*Figure 8: Task assignments for case 1*

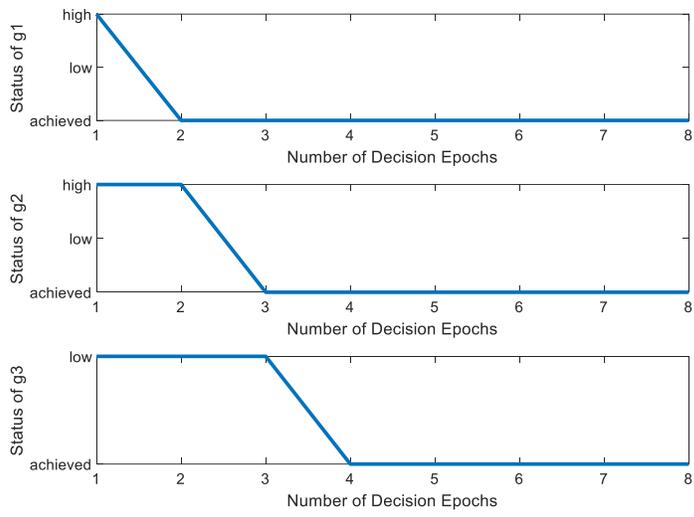

*Figure 9: Status of the goal flags for case 2*

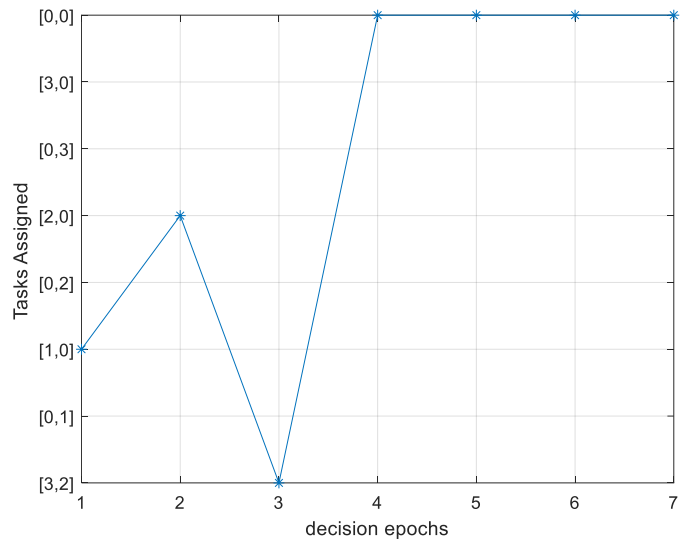

*Figure 10: Task assignments for case 2*

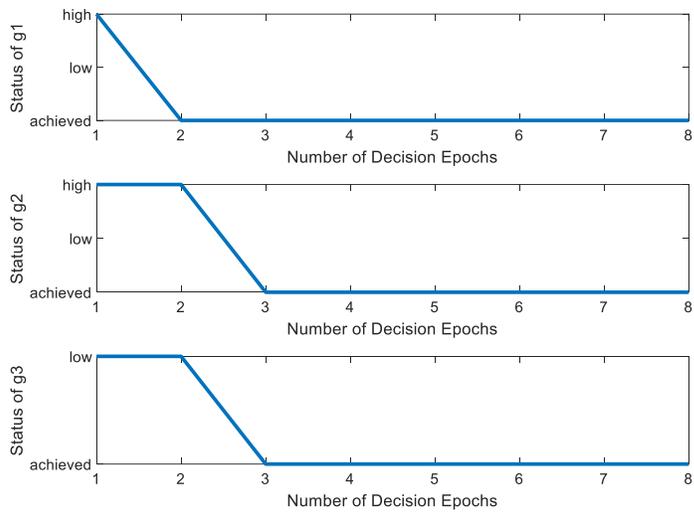

*Figure 11: Status of the goal flags for case 3*

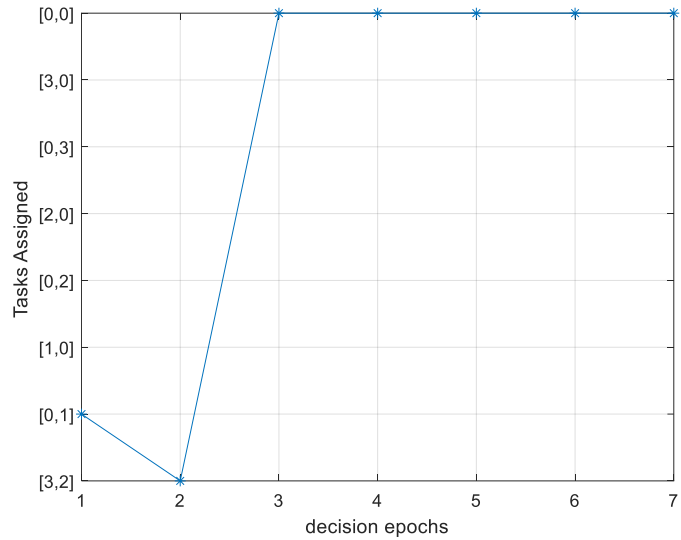

*Figure 12: Task assignments for case 3*

## 4.3 Policy trends

This subsection presents insights into the optimal task assignment policy and UAV top preferences. Specifically, we investigate the conditions on the state variables under which the optimal policy chooses certain assignments. The results for the UAV optimal bidding calculations are shown in Table 4. The variable values indicate that a UAV prefers a goal whenever it is within reach and that there is no fault present in the UAV. Battery recharging is preferred in states where the number of reachable (within the range of the UAV with the current SOC) goals is 2 or less. Furthermore, the average number of reachable goals in all the states where recharging has been the optimal decision is 0.7. The decision to get service is the most common decision in the policy because of our choice of low goal achievement probabilities under faulty conditions. Basically, the optimal policy reveals that the UAV's top priority is to get service as soon as any fault arises.

*Table 4: Conditions on state variables corresponding to various choices made by a UAV*

| UAV Top Preference | Conditions on State variables |
|---|---|
| Pursue Goal 1 | When goal 1 is within reach and there is no fault in the UAV |
| Pursue Goal 2 | When goal 2 is within reach and there is no fault in the UAV |
| Pursue Goal 3 | When goal 3 is within reach and there is no fault in the UAV |
| Recharge Battery | When the reachable goals are 2 or less (average 0.7 reachable goals) |
| Get Service | Whenever there is any fault, regardless of the other states |
| Do Nothing | When all of the goals are either out of reach or achieved |

The results from the task assignment policy have been presented in Table 5. The conditions on the state variables mentioned in the second column are self-explanatory. Two things are evident from the results. First, the decisions of the optimal policy are sensible and intuitive. Second, the MDP-based approach for dynamic task assignment offers an opportunity to carry out a detailed offline analysis regarding the consequences of various design choices in the model. Such a detailed exploration can help in finding out the key relationships between the optimal policy and the parameter values, such that the need to perform stochastic dynamic programming is reduced or omitted (in the ideal case). As a result, the scalability of the solution can be significantly enhanced. Therefore, such exploration is a promising avenue for future research.

*Table 5: Conditions on state variables corresponding to various choices in the optimal task assignment policy*

| Optimal Task Assignment | Conditions on State variables |
|---|---|
| [1, 2] | Both UAVs are available and fault-free. UAV 1 has goal 1 as its highest preference. Goal 3 has low priority or is achieved (or all three goals have high priority but goals 1 and 2 are preferred by the UAVs). |
| [2, 1] | Both UAVs are available and fault-free. UAV 1 has goal 2 as its highest preference. Goal 3 has low priority or is achieved (or all three goals have high priority but goals 1 and 2 are preferred by the UAVs). |
| [1, 3] | Both UAVs are available and fault-free. UAV 1 has goal 1 as its highest preference. Goal 2 has low priority or is achieved (or all three goals have high priority but goals 1 and 3 are preferred by the UAVs). |
| [3, 1] | Both UAVs are available and fault-free. UAV 1 has goal 3 as its highest preference. Goal 2 has low priority or is achieved (or all three goals have high priority but goals 1 and 3 are preferred by the UAVs). |
| [2, 3] | Both UAVs are available and fault-free. UAV 1 has goal 2 as its highest preference. Goal 1 has low priority or is achieved (or all three goals have high priority but goals 2 and 3 are preferred by the UAVs). |
| [3, 2] | Both UAVs are available and fault-free. UAV 1 has goal 3 as its highest preference. Goal 1 has low priority or is achieved (or all three goals have high priority but goals 2 and 3 are preferred by the UAVs). |
| [0, 1] | Goal 1 has high priority, and UAV 2 is available. UAV 1 is either unavailable, has developed a severe fault, or cannot reach goal 1 (due to low battery). |
| [1, 0] | Goal 1 has high priority, and UAV 1 is available. UAV 2 is either unavailable, has developed a severe fault, or cannot reach goal 1 (due to low battery). |
| [0, 2] | Goal 2 has high priority, and UAV 2 is available. UAV 1 is either unavailable, has developed a severe fault, or cannot reach goal 2 (due to low battery). |
| [2, 0] | Goal 2 has high priority, and UAV 1 is available. UAV 2 is either unavailable, has developed a severe fault, or cannot reach goal 2 (due to low battery). |
| [0, 3] | Goal 3 has high priority, and UAV 2 is available. UAV 1 is either unavailable, has developed a severe fault or cannot reach goal 3 (due to low battery). |
| [3, 0] | Goal 3 has high priority, and UAV 1 is available. UAV 2 is either unavailable, has developed a severe fault or cannot reach goal 3 (due to low battery). |

## 4.4 Comparison with existing approaches

To the best of our knowledge, there is no existing approach for dynamic task assignment for UAVs that can be compared directly with the proposed approach. The main reason for this gap is that all approaches differ in the features or aspects of the problem that they focus on. Some approaches are better in terms of collision avoidance, some are best in terms of energy efficiency, and some approaches accommodate faults. Therefore, in this section, we provide a qualitative comparison of our proposed approach with the existing ones. As discussed earlier, the major innovation in our proposed approach is the incorporation of sensing and actuation faults in the UAVs in terms of controllability and observability. Another innovation is the incorporation of UAV ranges (calculated from the available SOC information) with respect to the search goals in our proposed framework. Finally, the incorporation of serviceability and recharging into decision-making is also a new addition to the literature on UAV cooperative task assignment.

Table 6 provides a feature-based comparison of our proposed approach with six different types of approaches in the literature. The results in Table 6 confirm that our proposed approach is unique in terms of incorporating of faults. Also, the number of features included in our proposed approach is not available in any one of the existing approaches in combined form.

*Table 6: Feature-based analytical comparison of the proposed approach with some selected existing approaches*

| Task Assignment Approach | Incorporation of fault tolerance | Modeling of uncertainty in the problem | Optimization | Incorporation of battery SOC or range of UAV | Use of hierarchical approach |
|---|---|---|---|---|---|
| The proposed framework | Yes, using the system properties of controllability and observability. | Yes, the uncertainties are modeled directly in state transition probabilities | Yes, the decisions are optimized with respect to expected discounted cost function | Yes, the SOC is incorporated in terms of UAV ranges with respect to the search goals | Yes, the two-level hierarchical approach has been used |
| References [26][27] | No | Yes, the uncertainty is modeled in state transition probabilities | Yes, the decisions are optimized with respect to expected discounted cost function | Yes, the SOC level is modeled directly without calculation of the range | Yes, the two-level hierarchical approach has been used |
| Reference [16] | No faults considered but overall loss of UAV is considered | Possibility of UAV loss is considered but is not quantified | Machine learning is used to improve the solution | No | Yes, two-level approach has been used |
| Reference [17] | No | No | Yes, the decisions are optimized with respect to a cost function | No | Although not explicitly, but implicitly, the approach involves two levels of decision-making |
| Reference [20] | No | Yes, the uncertainty is considered in terms of the obstacles | The algorithm tries to minimize the chances of collision | No | No |
| Reference [21] | No | Uncertainty modeled in the collision avoidance part | Uses grey wolf optimization | No | Not hierarchical but the decision- |
| | | | | | making is distributed |
| Reference [22] | Actuator faults are considered in terms of actuator effectiveness | No | Yes, using decentralized Hungarian algorithm | No | Decentralized control is used. Also, the collision avoidance and navigation are performed separately |

# 5 Conclusions and Future Work

This paper provided an in-depth examination of a novel two-level decision-making framework designed for the dynamic task assignment of UAVs in a cooperative search scenario. This model differs from others in that it considers the post-fault capabilities inherent in low-level feedback control systems. Furthermore, by seamlessly integrating fault tolerance, uncertainty modeling, battery status, and stochastic optimization into a single, cohesive hierarchical structure, this strategy is intended to be comprehensive.

Undoubtedly, the successful implementation of the proposed approach depends on the selection of model parameters. Some parameters can be figured out from the operational and maintenance costs of UAVs, but others need to be figured out through experience or empirical data. For example, the probabilities of faults and changes in user-defined goal priorities are based on conditional probabilities.

Illustrated through a numerical simulation-based case study, the paper adeptly demonstrates the practical application of the proposed framework. It not only describes the execution of the framework, but also the consequences of parameter choices. A key feature of the model is its resilience in demanding scenarios, including instances of faults or diminished battery states of charge.

A major avenue for further investigation stemming from the introduced framework pertains to the exploration of scalability and reduction in the computational complexity of the solution. A crucial aspect in achieving heightened scalability involves an in-depth analysis of the intricate connection between the design parameters and the optimal policy. The seed for such research is provided in the case study discussed in the paper.

**Acknowledgement**

The authors would like to acknowledge the support provided by the Deanship of Research Oversight and Coordination (DROC) at King Fahd University of Petroleum & Minerals (KFUPM) for funding this work through project No. EC221013.